\documentclass[10pt,conference]{IEEEtran}
\IEEEoverridecommandlockouts
\usepackage{cite}
\usepackage{amsmath,amssymb,amsfonts}
\usepackage{algorithmic}
\usepackage{graphicx}
\usepackage{textcomp}
\usepackage{xcolor}
\usepackage[caption=false,font=footnotesize]{subfig}
\usepackage{comment}

\IEEEpubid{\begin{minipage}{\textwidth}
    \vspace{2cm} 
    \centering
    \fbox{\parbox{0.95\textwidth}{ 
        \footnotesize
        \copyright~2025 IEEE. Personal use of this material is permitted. Permission from IEEE must be obtained for all other uses, in any current or future media, including
 reprinting/republishing this material for advertising or promotional purposes, creating new collective works, for resale or redistribution to servers or lists, or
 reuse of any copyrighted component of this work in other works.
    }}
\end{minipage}}

\begin{document}

\title{\vspace{-1cm} 
\small This paper has been accepted for publication in the 2025 IEEE Wireless Communications and Networking Conference (WCNC).
\vspace{0.5cm} \\  
\huge Reliability Modeling for Beyond-5G Mission Critical Networks Using Effective Capacity}

\author{Anudeep Karnam, Jobish John, Kishor C. Joshi, George Exarchakos, Sonia Heemstra de Groot, Ignas Niemegeers\\
Center for Wireless Technology Eindhoven, Eindhoven University of Technology, The Netherlands}

\maketitle

\IEEEpubidadjcol 

\begin{abstract}
Accurate reliability modeling for ultra-reliable low latency communication (URLLC) and hyper-reliable low latency communication (HRLLC) networks is challenging due to the complex interactions between  network layers required  to meet stringent requirements. In this paper, we propose such a model. We consider the acknowledged mode of the radio link control (RLC) layer, utilizing separate buffers for transmissions and retransmissions, along with the behavior of  physical channels. Our approach leverages the effective capacity (EC) framework, which quantifies the maximum constant arrival rate a time-varying wireless channel can support while meeting statistical quality of service (QoS) constraints. We derive a reliability model that incorporates  delay violations, various latency components, and multiple transmission attempts. Our method identifies optimal operating conditions that satisfy URLLC/HRLLC constraints while maintaining near-optimal EC, ensuring the system can handle  peak traffic with a guaranteed QoS. Our model reveals critical  trade-offs between EC and reliability across various use cases, providing guidance for URLLC/HRLLC network design for service providers and system designers.
\end{abstract}
\section{Introduction}
Beyond-5G/6G networks are expected to support a wide variety of applications in different verticals by offering diverse services, including URLLC/HRLLC~\cite{ITU-R2030}. These services are essential for emerging applications that demand high reliability and deterministic wireless data transmissions, such as industrial automation, intelligent transportation systems, holographic telepresence, and tele-surgery~\cite{Tao2023}. Achieving extreme reliability with ultra-low latency, presents significant challenges across multiple layers of the mobile network stack and calls for a highly optimised network design, accounting for the interactions between them~\cite{Siddiqi2019}.

Physical channels are crucial in enabling mission-critical applications since they deal with several environment-dependent factors such as propagation losses, multi-path and fading effects~\cite{TKL}. The medium access control (MAC) layer challenges include real-time resource allocations, specifically, scheduling packets from multiple radio link control (RLC) queues based on channel conditions, which can introduce variability in latency~\cite{Bennis}. Addressing these challenges requires a reliability model that accounts for multi-layer interactions within the mobile network stack. We propose such a model that unifies the 5G link (RLC and MAC) and physical (PHY) layers, to aid  URLLC network design for mission-critical applications.

 In~\cite{Onel}, the authors provide an overview of statistical tools, including tail distribution approximations,  queuing and extreme value theory, and data-driven approaches highlighting their relevance to URLLC network design. One such statistical tool is the effective capacity (EC) \cite{Wu2003}, the dual of effective bandwidth~\cite{EB}, which quantifies the maximum constant arrival rate a communication system can support while satisfying quality of service (QoS) requirements.
The authors in~\cite{Musavian} optimize EC by jointly adapting power and datarate while satisfying packet error rate (PER) and delay constraints. However, they do not account for automatic repeat request (ARQ) or hybrid ARQ (HARQ) mechanisms. In contrast,~\cite{WCNC2016} incorporates HARQ in the EC model, revealing that retransmission overhead at high signal-to-noise ratio (SNR) can reduce EC, especially when using adaptive modulation and coding (AMC). The authors in~\cite{Muhammad2022} introduce mission effective capacity (MEC) and mission reliability (MR) metrics specifically for URLLC in industrial internet of things (IIoT), providing closed-form expressions for MR and mean time to first failure (MTTFF) utilizing the EC framework. ~\cite{HanRui2024} utilizes EC to analyze delay-sensitive systems and propose joint data rate and power allocation schemes that maximize EC.

However, existing works focus on models that consider only the unacknowledged mode (UM) of the RLC layer. This is a significant limitation because UM does not support ARQ retransmissions, whereas acknowledged mode (AM) allows feedback-based retransmissions, which are critical to meet strict reliability requirements. Existing reliability models do not account for separate physical channels for control and user plane information while accounting for unified link and physical layer characteristics in the reliability model. This is important since control plane information uses more robust modulation and coding schemes, resulting in  Correct decoding of downlink control information (DCI) that leads to accurate decoding of user plane information.  The  major contributions of this paper are:  
\begin{enumerate}
\item We extend the EC framework to a dual-buffer 5G RLC AM, deriving EC formulations for transmission (TX)  and retransmission (RETX) buffers.
\item We develop a reliability framework that considers multiple transmission attempts,  both user and control plane transmissions while accounting for deterministic and stochastic delays.
\item We quantify a specific range of operating conditions that satisfy URLLC constraints while ensuring EC remains close to its peak value, enabling the system to handle maximum arrival traffic while meeting QoS constraints. 
\item We provide practical insights into optimizing URLLC network performance by analyzing the trade-offs between EC and reliability across different use cases. 
\end{enumerate}
The rest of the paper is organized as follows: Sec.~\ref{sec_pblm_formulation} outlines the system model and problem formulation. Sec.~\ref{Sec_ReliabilityFramework} presents the proposed reliability framework. Sec.~\ref{sec_latency} presents the delay analysis. Sec.~\ref{Sec:Effective-Capacity-Reliability-Tradeoff} discusses the EC-reliability trade-offs. Sec.~\ref{Results} presents the results that demonstrate the optimal range of operating conditions and provide key insights for URLLC network design. Finally, Sec.~\ref{Conclusion} concludes the paper.


\section{System model and Problem Formulation}
\label{sec_pblm_formulation}
\subsection{System Model}
\label{sec_system_model}

We consider a system model comprising a single gNB and UE, focusing on the downlink scenario\footnote{The methodology can be applied to uplink transmissions by replacing the respective successful  transmission probabilities, as per the chosen uplink transmission scheme—whether grant-free or scheduling request-based.}. Our analysis focuses on a single QoS flow at the RLC layer, utilizing AM, leveraging ARQ retransmissions to meet the URLLC requirements. In AM mode, each QoS flow has a dedicated TX buffer and  RETX buffer. Once the packets arrive at the RLC layer, they are first stored in the TX buffer, undergo RLC procedures~\cite{RLC} and are sent to the RETX buffer before forwarding to the MAC layer. After MAC and PHY layer processing, packets are transmitted over the air interface through physical channels; physical downlink share channel (PDSCH)  for user/shared data transmissions, and physical downlink control channel (PDCCH) for downlink control information.

\begin{figure}[t]
\centering
\subfloat[gNB-UE Interactions]{\includegraphics[scale=0.38]{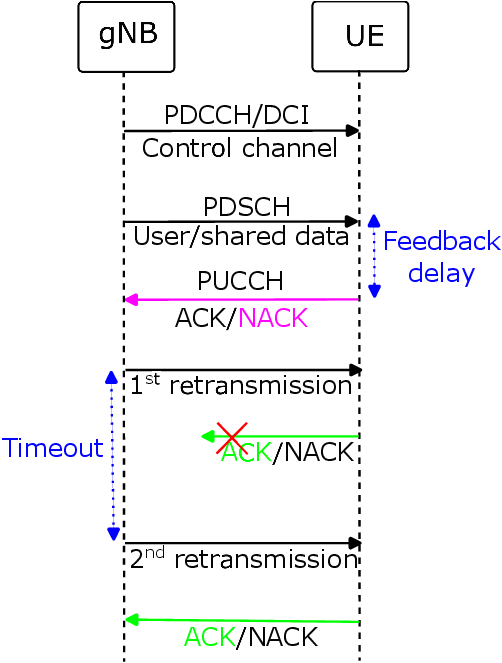}
\label{fig:1a}}
\subfloat[Tail regions of PDFs]{\includegraphics[scale=0.5]{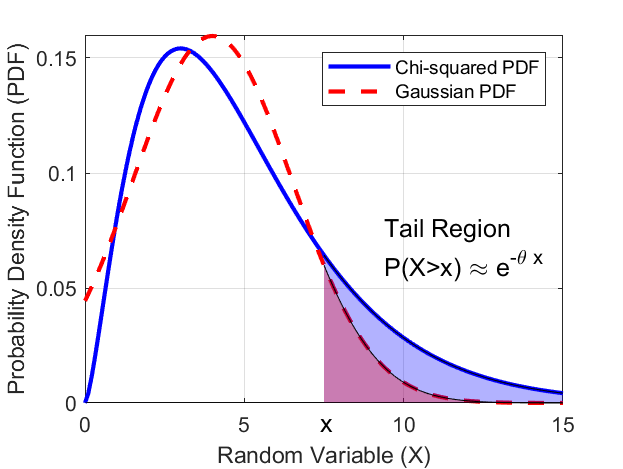}
\label{fig:1b}}
\hfil
\caption{System model and the exponential approximation of tail regions.}
\label{fig_one}
\end{figure}

Once UE receives a downlink packet, it decodes DCI, which
is further used to decode the user data on the PDSCH channel.
Following this decoding process, UE sends an acknowledgement (ACK) or no-acknowledgement (NACK)
to gNB through physical uplink control channel (PUCCH). When the gNB does not receive a timely acknowledgement, it initiates a retransmission after a predefined timeout interval $\mathcal{D}_{\textnormal{timeout}}$.
Fig.~\ref{fig:1a} shows the PHY layer interactions between the gNB and UE, highlighting the control/data flows, and retransmissions triggered by NACK or timeout.


\subsection{System Assumptions}
\label{SystemAssumptions}
\begin{enumerate}
    \item \textit{Retransmissions within the coherence time:} Retransmissions occur within the coherence time, ensuring static channel conditions. 
    Coherence time $T_{C}$ in seconds, can be approximated using the formula given in~\cite{Rappaport}, $T_C \approx \sqrt{\frac{9}{16\pi}}\frac{c}{vf_c}$, where $c$ is the velocity of light, $v$ is the relative velocity between the transmitter and receiver, and $f_c$ is the carrier frequency. For low-mobility applications with constant speed of $v=3  m/s$ and  $f_c = 2.4 GHz$, the coherence time is approximately $T_C = 17.6 ms$. In high-mobility cases the coherence time significantly reduces, making the assumption less applicable.
    While the assumption is reasonable for URLLC scenarios with limited mobility, high-mobility environments may require additional modeling to account for Doppler effects and rapid channel variations.
    \item \textit{Excludes soft-combining}: We focus on ARQ mechanisms at the RLC layer and excludes lower layer soft-combining techniques. 
    This simplification avoids the complexity of combining procedures and provides a conservative reliability estimate, acknowledging a slight underestimation compared to real-world scenarios.
\end{enumerate}
\subsection{Problem Formulation}

We utilize the large deviations principle (LDP)~\cite{Touchette2011ABI} to characterize the probability of rare events in any probability density function (PDF). The tail behavior of any PDF of a random variable (RV) \( X \) exceeding a given threshold \( x \), can be approximated as  $P(X > x) \approx e^{-\theta \cdot x}$, where \( \theta \) is the decay rate as shown in Fig. \ref{fig:1b}, the area under the curve defined by \( P(X > x) \) quantifies the violation probability. A larger \( \theta \) corresponds to a steeper decay (shown in red), indicating a smaller area and  lower violation probability, while a  smaller \( \theta \) implies a slower decay (shown in blue), resulting in higher  violation probability. 

In delay-sensitive communications, $\theta$ corresponds to the delay exponent, where higher values of $\theta$ represent a system with stricter delay constraints, and lower values indicate a system with longer delay tolerance.
We derive EC and reliability, denoted as $R(\theta,N)$ as functions of $\theta$, and number of transmission attempts $N$. To ensure that the system can handle peak arrival traffic, meeting the QoS constraints such as reliability and latency requirements, we define an optimization problem as the maximization of EC$(\theta,N)$  subject to the constraint that $R(\theta,N)$ exceeds a  threshold \(R_{\text{th}}\) as given in~(\ref{optimization}).

\begin{equation}
\label{optimization}
\begin{aligned}
    & \max_{\theta} \; \text{EC}(\theta, N) \\
    & \text{s.t.} \; R(\theta, N) \geq R_{\text{th}}.
\end{aligned}
\end{equation}

\section{Proposed Reliability Framework}
\label{Sec_ReliabilityFramework}
Our reliability framework captures physical channel transmissions and retransmissions due to timeouts/NACK, while accounting for delay violations. The delay violation probability (DVP) is defined as the likelihood that the total delay \(\mathcal{D}\) (from the RLC ingress at the gNB to the RLC egress at the UE), which includes both stochastic and deterministic delays  $\mathcal{D}_{\text{det}}$, exceeds a specified bound \(\mathcal{D}_{\text{max}}\) that is application dependent. If the queuing delay remains below \(\mathcal{D}_{\text{max}} - \mathcal{D}_{\text{det}}\), the total latency $\mathcal{D}$ is within acceptable bound $\mathcal{D}_{max}$ which  is used in our reliability formulations.


To formulate the reliability model for \( N \) transmission attempts, we analyze the reliability for $N=1, 2$, denoted as $R(N=1)$, and $R(N=2)$ respectively, followed by extending it to a generic model, denoted by $R(N)$. Let the probability of successful transmissions for PDCCH, PDSCH, and PUCCH be \( p_{1}, p_{2} \), and \( p_{3} \), respectively. The probability that a packet is successfully decoded in the first attempt, \( P(N=1) \), is given by $p_1 \times p_2$. Considering only processing delays at the UE and gNB, since no timeouts or feedback delays are involved when $N=1$, DVP can be captured as \( P(\mathcal{D} > \mathcal{D}_{\text{max}} - \mathcal{D}_{\text{det}}) \) and hence,
\begin{equation}
\label{eq_reliability_model_1}
R(N = 1)  = (1 - P(\mathcal{D} > \mathcal{D}_{\text{max}} - \mathcal{D}_{\text{det}})) \times P(N = 1)        
\end{equation}
Multiple transmission scenarios can arise either due to a timeout event or  the reception of a NACK. A timeout triggered retransmission event occurs if; (1) PDCCH fails, leading to no feedback; or (2) PDCCH succeeds, PDSCH fails, and NACK is lost. Thus the probability of timeout is $
P(\text{timeout}) = (1 - p_1) + p_1 (1 - p_2)(1 - p_3).$
Retransmission due to NACK occurs, when PDCCH is successfully decoded, but PDSCH data decoding fails, and the gNB receives a NACK, This probability is given by $P(\text{feed}) = p_1 (1 - p_2) p_3.$
The DVPs for the second transmission attempt, triggered by either timeout or NACK are \( P(\mathcal{D} > \mathcal{D}_{\text{max}} - 2 \mathcal{D}_{\text{det}} - \mathcal{D}_{\text{timeout}}) \) and \( P(\mathcal{D} > \mathcal{D}_{\text{max}} - 2 \mathcal{D}_{\text{det}} - \mathcal{D}_{\text{feed}}) \) respectively, where \( \mathcal{D}_{\text{feed}} \) is the feedback delay. The overall reliability for \( N = 2 \) is:
\begin{align}
& R(N = 2) = R(N = 1) + \nonumber \\
& \left(1 - P(\mathcal{D} > \mathcal{D}_{\text{max}} - 2  \mathcal{D}_{\text{det}} - \mathcal{D}_{\text{timeout}})\right)  P(\text{timeout}) \cdot  p_1 p_2 \nonumber \\
& +\left(1 - P(\mathcal{D} > \mathcal{D}_{\text{max}} - 2  \mathcal{D}_{\text{det}} - \mathcal{D}_{\text{feed}})\right)  P(\text{feed}) \cdot p_1 p_2 \label{eq_reliability_model_2}
\end{align}
We extend the above approach used to obtain (\ref{eq_reliability_model_1}) and (\ref{eq_reliability_model_2}) by considering the cumulative effects of both NACK feedback and timeout triggered multiple transmission attempts. The total deterministic delay, \(\mathcal{D}_{\text{det}}\), increases linearly with the number of transmission attempts, and the delays due to feedback and timeout events also finds its position in the reliability model. For the $N^{th}$ transmission attempt, we consider all the possible combinations of feedback and timeout events that could occur during the previous transmission attempts. We use the binomial coefficient \(\binom{N-1}{k}\) to represent the number of ways \(k\) timeout events and \(N-1-k\) feedback events can occur. Each term in the summation (refer (\ref{eq_reliability_model_N})) corresponds to the probability of having exactly \(k\) timeouts and \(N-1-k\) feedback events, along with their respective delay contributions. Thus the precise reliability model that accommodates $N$ transmission attempts incorporating latency contraints, NACK feedback and timeout-triggered transmissions, is given by:
\begin{align}
\label{eq_reliability_model_N}
&R(N) = R(N-1) +\sum_{k=0}^{N-1} \bigg[ \bigg( 1 - P\big(\mathcal{D} > \mathcal{D}_{\textnormal{max}} - N\mathcal{D}_{\textnormal{det}} \nonumber \\
&\qquad- k\mathcal{D}_{\textnormal{timeout}} - (N-1-k)\mathcal{D}_{\textnormal{feed}}\big) \bigg) \nonumber \\
& \times \binom{N-1}{k} P(\textnormal{feed})^{N-1-k}  P(\textnormal{timeout})^{k}\times p_1 p_2 \bigg]
\end{align}

\section{Delays and Delay Violations}
\label{sec_latency}
\begin{figure}[t]
\centering
\includegraphics[scale=0.5]{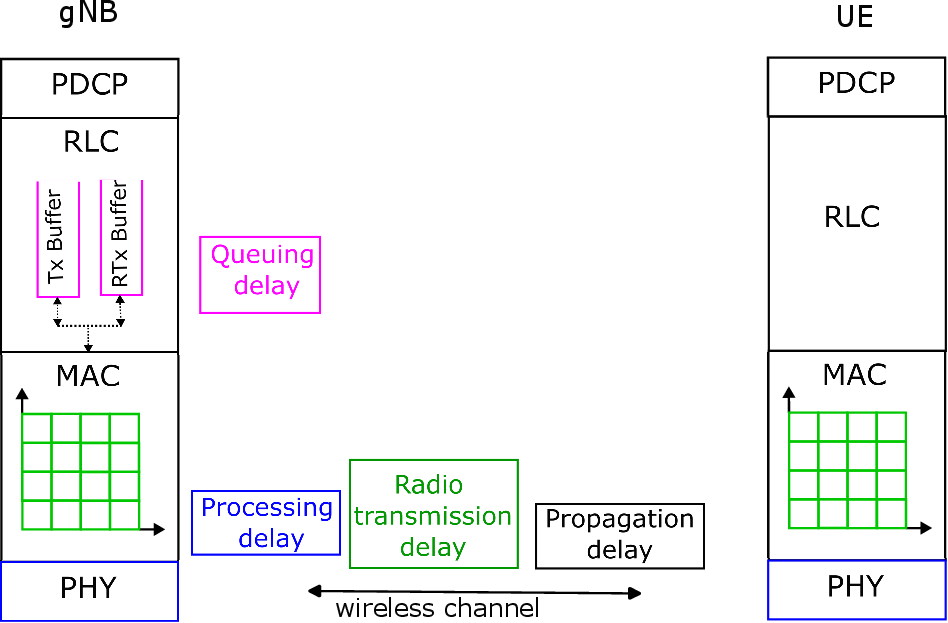}
\caption{Delay components between gNB and UE}
\label{Fig_latency}
\end{figure}

This section analyses user plane latency, which includes deterministic and stochastic delays as shown in Fig.~\ref{Fig_latency}.
\subsection{Deterministic Delays}
\subsubsection{Processing Delays}
Processing delays occur at both gNB and UE due to tasks such as encoding, modulation, equalizing, and decoding~\cite{Chapter12}. The processing delay at the gNB, assumed constant for both transmission and retransmissions, is denoted as \(\mathcal{D}_{\textnormal{gNB}}\) while the processing delay at the UE, from  reception to the ACK/NACK decision, is denoted as \(\mathcal{D}_{\textnormal{UE}}\).
\subsubsection{Feedback and Timeout Delays}
Feedback delay (\(\mathcal{D}_{\textnormal{feed}}\)) includes the transmission time of feedback from the UE to the gNB and the processing time at the gNB to interpret it. Timeout delay (\(\mathcal{D}_{\textnormal{timeout}}\)) occurs when the gNB does not receive an ACK/NACK within the expected time, including the waiting period before confirming feedback failure. For URLLC/HRLLC  applications, typically, propagation distances are small. Hence, we do not account propagation delays in our analysis. 
\subsubsection{Transmission Delays in the Finite Blocklength Regime}
To model the time and frequency resources required for URLLC in the short blocklength regime, we use the normal approximation from \cite{Polyanskiy} given by (\ref{Shannon}), as Shannon's limit is not suitable for URLLC's shorter blocks. In this regime, with \( r \) channel uses (Hz·s) and \( L \) information bits, the achievable channel service rate \( S \) is \( S = L/r \) (bits per channel use (bpcu) or bits/Hz/s). For an AWGN channel with decoding error probability \( \epsilon \), SNR, and \( L \) bits in \( r \) channel uses, the achievable rate is:
\begin{equation}
\label{Shannon}
S = \frac{L}{r} \approx C - Q^{-1}(\epsilon) \sqrt{\frac{V}{r}}
\end{equation}
where \( C = \log_2(1 + \text{SNR}) \) is the Shannon capacity limit, \( V = (\log_2 e)^2 \left( 1 - \frac{1}{(1 + \text{SNR})^2} \right) \) is the channel dispersion, and \( Q^{-1}(\cdot) \) is the inverse Q-function.
Using (\ref{Shannon}), the number of channel uses \( r \) needed to transmit \( L \) bits with decoding error probability \( \epsilon \) can be approximated as:
\begin{equation}
\label{Eq_r}
\begin{aligned}
r \approx \frac{L}{C} + \frac{\left( Q^{-1}(\epsilon) \right)^2 V}{2 \left( C \right)^2} \left( 1 + \sqrt{1 + \frac{4 L C}{V \left( Q^{-1}(\epsilon) \right)^2}} \right)
\end{aligned}    
\end{equation}
Let \( B \) be the bandwidth allocated to the UE and \( T \) be the corresponding time resources. Then, \( r \) can be calculated as \( r = B \times T \). Assuming fixed bandwidth \( B \), the transmission time \( T \) needed to transmit \( L \) bits are given by \( T = \frac{r}{B} \). This \( T \) is deterministic under constant SNR assumption, which holds in high coherence time scenarios typical of URLLC with low mobility.
The total deterministic latency, \(\mathcal{D}_{\textnormal{Det}}\), accounts for multiple transmissions, including processing, feedback, and timeout delays. For \(N\) transmission attempts with \(a_{1}\) feedback events and \(a_{2}\) timeout events, the total deterministic delay is:
\begin{equation}
\label{eq_Ddet}
\begin{aligned}
\mathcal{D}_{\textnormal{Det}} = N(\mathcal{D}_{\textnormal{gNB}} + \mathcal{D}_{\textnormal{UE}} + T) + a_{1}\mathcal{D}_{\textnormal{feed}} + a_{2}\mathcal{D}_{\textnormal{timeout}} 
\end{aligned}
\end{equation}

\subsection{Stochastic delay analysis in 5G RLC buffers}
\label{nondeterministicdelays}
Here we discuss the dynamics  of TX and RETX buffers and establish conditions that prevent indefinite queue growth, which leads to excessive delays. We also derive the EC for these buffers, and provide the queuing DVP by leveraging LDP.    
The queue dynamics of TX  buffer can be captured as:
\begin{equation}
\label{txqlength}
Q_{\text{TX}}(t) = Q_{\text{TX}}(0) + A_{\text{TX}}(t) - S_{\text{TX}}(t)
\end{equation}
where \( A_{\text{TX}}(t) \)  and \( S_{\text{TX}}(t) \) denote the packet arrival and service process, respectively, and \(Q_{\text{TX}}(t)\)  represents queue length at time~$t$. Stability is achieved if  \( A_{\text{TX}}(t) \) and \( S_{\text{TX}}(t) \) are stationary and ergodic \cite{Wu2003} with average  arrival rate~$\lambda_{\text{TX}}$, less than average service rate  $\mu_{\text{TX}}$, preventing indefinite queue growth.
Similarly, the queue dynamics for the RETX buffer is given by:
\begin{equation}
\label{retxqlength}
Q_{\text{RETX}}(t) = Q_{\text{RETX}}(0) + A_{\text{TX}}(t) - S_{\text{RETX}}(t)
\end{equation}
Since all packets arrive at RETX buffer in the RLC AM mode, we have \(\lambda_{\text{RETX}} = \lambda_{\text{TX}}\). Packets are removed from the RETX buffer upon ACK reception (denoted by the event  $\mathcal{E}(\text{ACK})$), or when the retransmission limit is reached $(N_{\text{max}})$, or if delay constraint \(\mathcal{D}_{\text{max}}\) is violated. If NACK is received, packets are scheduled for retransmission. 
Thus, the effective service rate of the RETX buffer considering multiple attempts is given by: $\mu_{\text{RETX}} =  \alpha \cdot \mathbb{E}[S_{\text{RETX}}(t)]$
where, $\alpha = P( {\mathcal{E}(\text{ACK}}) \cup N = N_{\text{max}} \cup \mathcal{D} > \mathcal{D}_{\text{max}}    )$. The stability of the RETX buffer requires that the rate at which packets are retained due to NACKs does not exceed the effective service rate; $\lambda_{\text{RETX}} < \mu_{\text{RETX}}$  ensuring that the buffer maintains a bounded queue length over time. 
In 5G RLC, the TX and RETX buffers form a unified system, and \(\theta\) serves as a delay exponent that captures the effect of packet transmissions from both the TX and RETX buffers. 

\subsubsection{EC for TX Buffer}
The EC is defined as $\text{EC}(\theta) = \frac{-1}{\theta}\text{log}\mathbb{E}[e^{-\theta S}]$~\cite{Wu2003}, where $S$ is the cumulative service rate given in (\ref{Shannon}).   For the TX buffer, the service process depends on two outcomes. First,  with probability \(p_1 p_2\), representing the successful decoding of both control  and user plane  data, the service rate is \(S\). Alternatively, if transmission fails  with probability \(1 - p_1 p_2\),  the  service rate is zero. Thus, the expected value of the moment generating function (MGF) is $\mathbb{E}\left[e^{-\theta S}\right] = (1 - p_1 p_2) + p_1 p_2 e^{-\theta S}$. Therefore, the EC for TX Buffer can be represented as:
\begin{equation}
\label{ctrans}
    \text{EC}(\theta, N=1) = -\frac{1}{\theta} \log\left[(1 - p_1 p_2) + p_1 p_2 e^{-\theta S}\right]
\end{equation} 
\subsubsection{EC for RETX Buffer}
The RETX buffer EC formulation captures the impact of  retransmissions as shown in (\ref{Cretrans}). 
The  MGF of the service process  depends on each transmission attempt's success. Unlike the TX buffer, if a packet fails in its first transmission attempt, it is retransmitted until it succeeds or reaches the maximum number of allowed transmission attempts. Thus, the summation term in (\ref{Cretrans}) represents the cumulative contribution of the service process over all \( N \) retransmission attempts. The term \( (1 - p_1 p_2)^N \) captures the scenario where the packet has undergone \( N \) attempts and is discarded after reaching the maximum transmission limit without success. Therefore, the expected value of the MGF over \( N \) transmission attempts is: 
\begin{align}
\label{Eq_Cretrans_MGF}
\mathbb{E}\left[ e^{-\theta S_{\text{RETX}}} \right] = (1 - p_1 p_2)^N + \sum_{k=0}^{N-1} (1 - p_1 p_2)^k _1 p_2 e^{-\theta (N - k)S}
\end{align}
 The EC for the RETX buffer is given by:
\begin{align}
\label{Cretrans}
&\textnormal{EC}( \theta,N) =-\frac{1}{N \theta} \log\Big[(1 - p_1p_2)^N + p_1p_2e^{-\theta S} \nonumber \\
&\qquad + \sum_{k=1}^{N-1} (1 - p_1p_2)^k p_1 p_2 e^{-\theta (N - k) S} \Big] 
\end{align}
\subsubsection{Queueing Delay Violation Probability}
The DVP is crucial in determining whether the total system delay \(\mathcal{D}\) exceeds the maximum allowable delay \(D_{\text{max}}\). Using the LDP, the queueing DVP can be approximated as $P(\mathcal{D} > \mathcal{D}_{\text{max}}) \approx \eta e^{-\theta S}$ \cite{Wu2003}, where \(\eta\) is the probability of the buffer being non-empty, $\approx 1$ \cite{HanRui2024}. To extend this model for multiple transmission attempts and incorporate deterministic delays, we refine the DVP expression using the total deterministic delay \(\mathcal{D}_{\text{det}}\)  derived  in  (\ref{eq_Ddet}). For the total latency \(\mathcal{D}\) to remain within acceptable bounds \(D_{\text{max}}\), the queuing delay must satisfy $\mathcal{D}_{\text{stoch}} < D_{\text{max}} - D_{\text{det}}$. Thus, the refined  DVP  accounting for both stochastic delays and deterministic delays is:
\begin{equation}
P(\mathcal{D} > D_{\text{max}} - D_{\text{det}}) \approx e^{-\theta (D_{\text{max}} - D_{\text{det}})},
\end{equation}

\section{Effective Capacity-Reliability Tradeoff}
\label{Sec:Effective-Capacity-Reliability-Tradeoff}
We aim to solve the optimization problem formulated in (\ref{optimization}), to outline the trade-off between EC and reliability.

We aim to maximize $EC(\theta, N$ (\ref{optimization}), to outline the trade-off between EC and reliability.
We consider \( h(\theta) = \log(g(\theta)) \), where \(g(\theta)\) is the expression within the logarithm in (\ref{Cretrans})  and examine the nature of $h(\theta)$ by computing its second derivative, given as: $h''(\theta) = \frac{g''(\theta) g(\theta) - (g'(\theta))^2}{(g(\theta))^2}.$
Applying Cauchy-Schwarz inequality, we obtain $(g'(\theta))^2 \leq g''(\theta) g(\theta),$
confirming that \( h''(\theta) \geq 0 \), hence \( h(\theta) = \log(g(\theta)) \) is convex. 

To maximize \(\text{EC}(\theta)\), it is essential that the function is concave. We consider $\Phi(\theta) = -N\cdot\text{EC}(\theta,N) = \frac{h(\theta)}{\theta}$, where \(h(\theta)\) is convex. To prove   \(\Phi(\theta)\) is convex, we  verify that for any \(\theta_1, \theta_2 > 0\) and \(\lambda \in [0, 1]\), the following inequality holds: 
\begin{equation}
\label{EC_concave_eq}
\Phi(\lambda \theta_1 + (1 - \lambda) \theta_2) \leq \lambda \Phi(\theta_1) + (1 - \lambda) \Phi(\theta_2)
\end{equation}
Since \(\Phi(\theta) = \frac{h(\theta)}{\theta}\) and \(h(\theta)\) is convex, we employ Jensen’s inequality to \(h(\theta)\) and  dividing both sides by \(\lambda \theta_1 + (1 - \lambda) \theta_2\),
\begin{equation}
\frac{h(\lambda \theta_1 + (1 - \lambda) \theta_2)}{\lambda \theta_1 + (1 - \lambda) \theta_2} \leq \frac{\lambda h(\theta_1) + (1 - \lambda) h(\theta_2)}{\lambda \theta_1 + (1 - \lambda) \theta_2}.
\end{equation}
Using the property that the weighted average of fractions is less than or equal to the fraction of weighted averages,
\begin{equation}
\label{EC_concave_eq}
\frac{\lambda h(\theta_1) + (1 - \lambda) h(\theta_2)}{\lambda \theta_1 + (1 - \lambda) \theta_2} \leq \lambda \frac{h(\theta_1)}{\theta_1} + (1 - \lambda) \frac{h(\theta_2)}{\theta_2}.
\end{equation}
This inequality proves that $\Phi$ is convex in $\theta$ and hence $\textnormal{EC}(\theta) = -\frac{\Phi(\theta)}{N}$ is concave.

We define the Lagrangian function: $\mathcal{L}(\theta, \lambda) = \text{EC}(\theta) - \lambda (R_{\text{th}} - R(\theta)),$ where \(\lambda \geq 0\) is the Lagrange multiplier and solve the optimizing problem  using the Karush-Kuhn-Tucker (KKT) conditions. The first condition is stationarity, requiring that the gradient of the Lagrangian with respect to \(\theta\) is zero: $\frac{\partial \mathcal{L}}{\partial \theta} = \frac{\partial \text{EC}(\theta)}{\partial \theta} + \lambda \frac{\partial R(\theta)}{\partial \theta} = 0,$
which balances the gradients of the Effective Capacity and Reliability function, scaled by \(\lambda\). The primal feasibility condition ensures that the reliability constraint \( R(\theta) \geq R_{\text{th}} \) is satisfied,  while dual feasibility condition imposes that \(\lambda \geq 0\). Complementary slackness requires $\lambda (R_{\text{th}} - R(\theta)) = 0$, indicating  the constraint is either active or inactive. If the constraint is inactive (\(R(\theta) > R_{\text{th}}\)), then \(\lambda = 0\), allowing unconstrained maximization of \(\text{EC}(\theta)\). Conversely, when the constraint is active (\(R(\theta) = R_{\text{th}}\)), \(\lambda > 0\), signifying that the reliability requirement directly impacts the optimization. Therefore, the optimal value of \(\theta\), denoted as $\theta_{\text{min}}$, occurs when the reliability constraint is tight, i.e., \(R(\theta_{\text{min}}) = R_{\text{th}}\),   $\theta_{\text{min}} = R^{-1}(R_{th})$. For all $\theta>\theta_{\text{min}}$, $R>R_{\text{th}}$ is met. However, as $\theta$ is increases EC will decrease  since EC is monotonically decreasing with $\theta$. Hence, an upper bound on $\theta$ is required.
We seek such a point \( \theta = \theta_{\text{max}} \) in the domain of \( \textnormal{EC}(\theta,N) \) that is near its supremum, \(\mathcal{M} = \sup_{\theta_{\text{max}} > 0} \textnormal{EC}(\theta_{\text{max}})\). Specifically, we aim to find \( \theta_{\text{max}} \in (0, \infty) \) such that \( \textnormal{EC}(\theta_{\text{max}}) \) is close to \(\mathcal{M} - \rho\),  where \(\rho > 0\) is a small positive constant that defines allowable degradation in EC. This proximity is characterized by minimizing the logarithmic distance:
\begin{equation}
\theta_{\text{max}} = \arg\min_{\theta > 0} \log \left( \left| \text{EC}(\theta) - (\mathcal{M} - \rho) \right| \right)
\end{equation}

The logarithmic transformation compresses variations, emphasizing points near the threshold \(\mathcal{M} - \rho\).  The gradient of the objective function \( \Psi(\theta) = \log |\text{EC}(\theta)- (\mathcal{M} - \rho)| \) is approximated using central finite differences:
\begin{equation}
\nabla \Psi(\theta) \approx \frac{\Psi(\theta + \delta) - \Psi(\theta - \delta)}{2\delta},
\end{equation} where \(\delta\) is a small perturbation ensuring stable numerical estimation. 
The operating region \([\theta_{\min}, \theta_{\max}]\) is derived to balance EC and reliability. \(\theta_{\min}\) is determined by optimizing \(\text{EC}(\theta, N)\) subject to the reliability constraint \(R(\theta, N) \geq R_{\text{th}}\), and \(\theta_{\max}\) is obtained using standard gradient-descent, minimizing the distance to the EC supremum (\(\mathcal{M} - \rho\))---beyond which EC starts to decay---ensuring that EC is within acceptable bounds.
\section{Results and Discussion}
\label{Results}

This section discusses how the proposed reliability model is used to optimize the trade-offs between the EC,  and reliability for URLLC/HRLLC applications. The proposed reliability model is independent of the methods used to obtain the probabilities \( p_1, p_2, \text{and } p_3 \). In our case, 5G PHY layer simulations are conducted under AWGN channel conditions, focusing on the finite blocklength regime to obtain these values. We consider a downlink scenario with 20 MHz bandwidth and 30 kHz subcarrier spacing, allocating 51 PRBs for both control and user data. Control traffic utilizes BPSK modulation and polar codes with list decoding, while data traffic uses 16QAM with LDPC coding, consistent with 3GPP standards. The deterministic latency $\mathcal{D}_{\text{Det}}$ is calculated by considering processing delays at the gNB. For numerology~1 (30 kHz subcarrier spacing), this delay amounts to 7 OFDM symbol duration (OS) \cite{Chapter12}. The transmission delay $T$ for short packets is given by $T = \frac{r}{B}$, where $r$ is computed using the finite blocklength approximation in (\ref{Eq_r}), assuming a 20 MHz bandwidth and a packet size of 32 bytes. At the UE, the PDSCH processing duration until ACK/NACK decision is considered as 4.5 OS duration \cite{Chapter12}.

\subsection{Effective Capacity and Reliability Trade-offs}
\label{Sec_EC-Rel-Tradeoffs}
\begin{figure}[htbp]
\centering
\hspace{-1.5 em}
\subfloat[EC($\theta,N$)]{\includegraphics[scale=0.41]{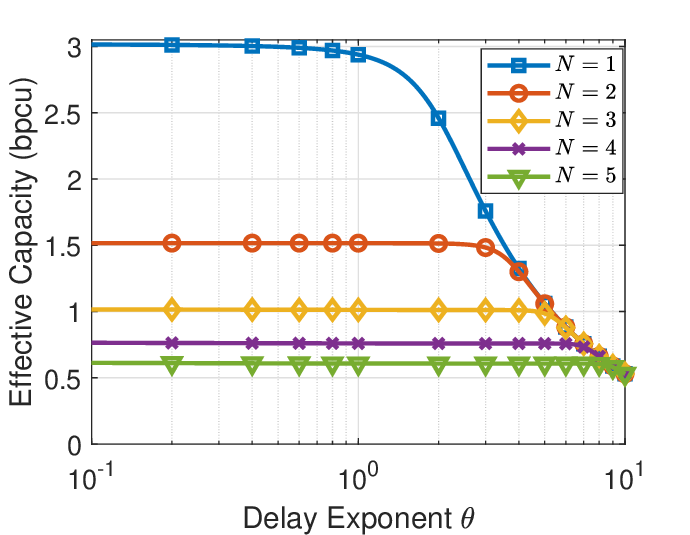}
\label{fig_EC_Rel_a}}
\hspace{-2.5em}
\hfil
\subfloat[R($\theta,N$)]{\includegraphics[scale=0.41]{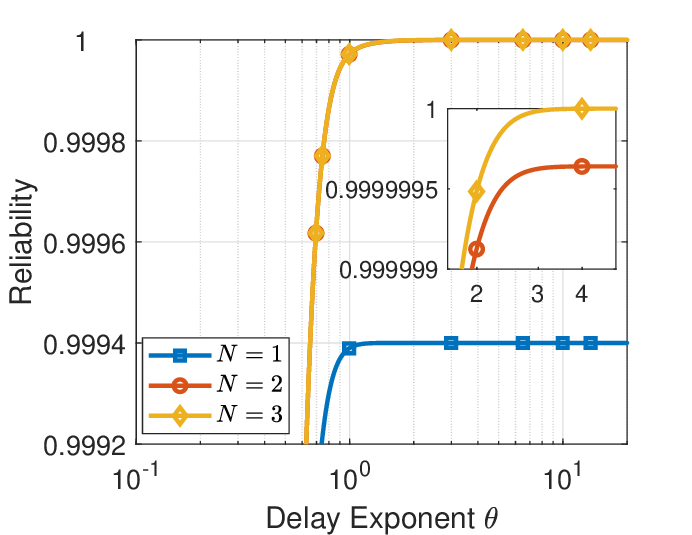}
\label{fig_EC_Rel_b}}
\hspace{-2.5em}
\hfil
\caption{Effective capacity and Reliability trade-offs}
\label{fig_EC_Rel}
\end{figure}

Fig. \ref{fig_EC_Rel_a}  illustrates the relationship between the delay exponent \(\theta\) and the EC for various $N$. As \(\theta\) increases, representing stricter  DVP QoS constraints, the EC exhibits a monotonically decreasing trend. For \(N = 1\), the EC remains relatively constant at around 3~bits/Hz/s for \(\theta \leq 1\), indicating that under relaxed DVP constraints, the EC saturates at the service rate. However, it drops sharply as QoS becomes more stringent. Specifically, when \(\theta\) approaches \(10^1\), the EC declines to approximately 0.5 bits/Hz/s, which reflects an 83\% reduction from its peak. This significant reduction highlights the network's diminished ability to handle incoming traffic under tighter QoS conditions. While a higher \(\theta\) indicates a lower DVP for a given \(D_{\text{max}}\), this improvement in DVP and hence reliability, comes at the expense of a notable reduction in EC. 
Fig. \ref{fig_EC_Rel_b}  illustrates the reliability performance \(R(\theta, N)\) for various transmission attempts \(N = 1, 2, 3\) revealing that reliability increases with retransmission attempts. For \(N = 1\), the reliability saturates at approximately 99.94\%, while for \(N = 2\), reliability reaches nearly 99.9999\%. The highlighted region in Fig. \ref{fig_EC_Rel_b} shows that reliability approaches ultra-high levels $(>99.9999\%)$, specifically for $N=2, 3$. However, the diminishing returns with additional attempts are evident,  as the reliability improvement from \(N = 2\) to \(N = 3\) is marginal ($\approx 0.0001\%$), suggesting that beyond a certain $N$, further retransmissions offer limited benefit in reliability. In such cases, employing other techniques, such as frequency or space diversity, may be beneficial compared to an increased number of retransmissions.
As $N$ increases, the EC further decreases, indicating that higher reliability comes at the cost of reduced traffic capacity that can meet the QoS requirements. This highlights the trade-off between reliability and EC, where additional retransmission attempts improve reliability but limit the amount of traffic the network can efficiently handle within the given QoS constraints.

\subsection{Optimal Operating Region}

\begin{figure}[h]
\centering
\includegraphics[scale = 0.45]{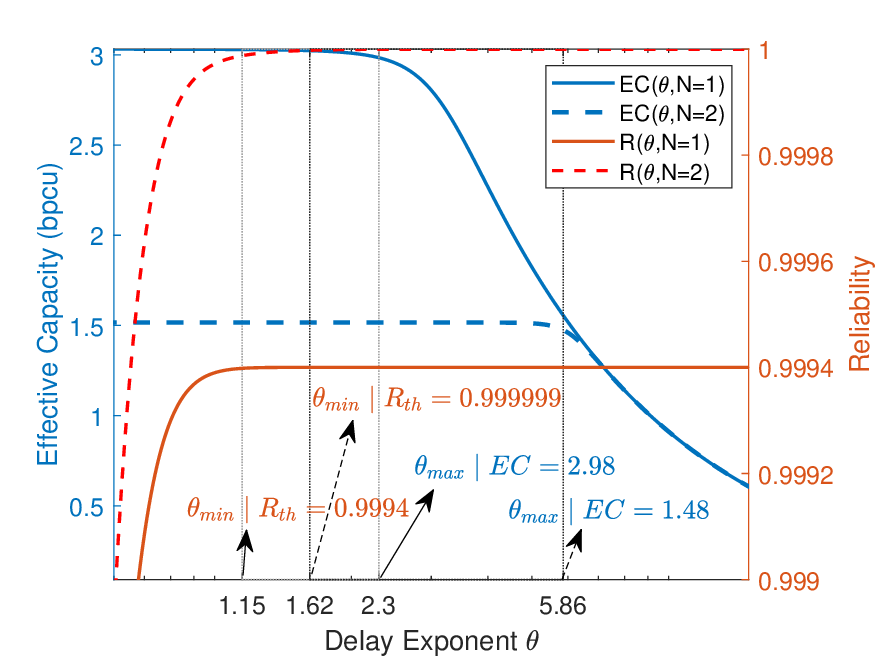}
\caption{Optimal range \([\theta_{\text{min}}, \theta_{\text{max}}]\) with \(R \geq R_{\text{th}}\)}
\label{OptimalRange}
\end{figure}

In this section, we determine the optimal range for the delay exponent \(\theta \in [\theta_{\text{min}},\theta_{\text{max}}]\). We  calculate \(\theta_{\text{max}}\) using the gradient descent as discussed in section \ref{Sec:Effective-Capacity-Reliability-Tradeoff}. \(\theta_{\text{min}}\) represents the minimum value of the delay exponent which ensures that the reliability requirement is met \((R = R_{\text{th}})\). Fig.  \ref{OptimalRange} illustrates the optimal region of operation is defined by \([\theta_{\text{min}}, \theta_{\text{max}}]\) for both \(N=1\) and \(N=2\). For all \(\theta \geq \theta_{\text{min}}\), the reliability condition \(R \geq R_{\text{th}}\) holds, ensuring the required system reliability. On the other hand, \(\theta_{\text{max}}\) is the point beyond which the effective capacity \(\text{EC}(\theta)\) starts to degrade. 
This range \([\theta_{\text{min}}, \theta_{\text{max}}]\), provides the optimal operating region such that the reliability constraints are satisfied with minimal degradation in \(\text{EC}(\theta, N)\). Beyond \(\theta_{\text{max}}\), EC declines, indicating suboptimal network performance. Thus, balancing \(\theta\) within this range ensures high system performance while maintaining the required reliability.

\subsection{Guidelines for URLLC Network Design}

This section provides practical guidelines for URLLC network design making use of the proposed reliability-EC model which can be used to analyze the trade-offs between EC, and reliability. Table \ref{table2} summarizes key use cases with respective reliability requirements. For ultra-low latency applications such as telepresence, with a 1 ms delay requirement, we can achieve 99.94\% reliability with a single transmission based on our analysis. However, this falls short of the 99.999\% reliability requirement. As clearly visible from the table, accommodating two transmissions helps to achieve a maximum of 99.9999\% reliability, albeit at the cost of reduced EC (down to 1.48 bpcu). This is feasible for uses cases such as remote control, energy distribution systems etc., which has a relaxed delay requirement (5ms). However if we need to satisfy strict delay requirements (for e.g., 1ms in case of telepresence use case) with ultra high reliability $(>99.9999\%)$ additional reliability enhancement techniques are essential. Higher numerology is one of the alternative techniques that helps to accommodate more retransmissions, while others include  techniques such as frequency and  diversity, and multi-connectivity. 
Conversely, applications such as  navigation  and flight control systems within wireless avionics intra-communications (WAIC\footnote{WAIC refers to communication systems designed exclusively for onboard data communications within an aircraft. These systems are not intended for air-to-air, air-to-ground, or ground-to-air communications. Both the transmitter and receiver are located within the aircraft, ensuring a minimal relative velocity between them. This satisfies the assumption of low mobility described in Section \ref{SystemAssumptions}}) that can tolerate higher latencies but demand ultra-high reliability $(\geq 99.99999\%)$ benefit from multiple transmission attempts.  Our results show that allowing up to three transmission attempts ($N=3$) and operating $\theta$ within the range of [2.45, 6.80] can meet the  reliability requirement of 99.99999\% with an EC of 1.01 bpcu. For even more demanding applications such as flight controls requiring 99.9999999\%  reliability, operating within the extended range [3.85, 8.65] can achieve this target with an EC of 1.0 bpcu. Notably, the reduction in EC is minimal compared to the significant improvements in reliability. Network operators must recognize that such extreme scenarios demand compromises on EC which is unavoidable. 
\begin{table}[h!]
\centering
\caption{Use Case Analysis: Operating Conditions for URLLC}
\setlength{\tabcolsep}{3pt} 
\renewcommand{\arraystretch}{1.1} 
\scriptsize 
\begin{tabular}{|p{2.35 cm}|p{0.6cm}|c|c|c|p{1.1cm}|p{0.61cm}|}
\hline
\textbf{Use Case} & \textbf{$D_{\text{max}}$ (ms)} & \textbf{$R_{\text{th}}$ (\%)} & \textbf{N} & \textbf{[\(\theta_{\text{min}}, \theta_{\text{max}}\)]} & \textbf{Achievable} \newline \textbf{Reliability} & \textbf{EC} \textbf{(bpcu)} \\
\hline
Telepresence & 1 & 99.999 & 1 & [1.15, 2.30] & 99.94 & 2.98 \\
\hline
 Discrete automation \newline User Experience & 1 & 99.9999 & 1 & [1.15, 2.30] & 99.94 & 2.98 \\
\hline
 Electricity distribution \newline high voltage &5 & 99.9999 & 2 & [1.62, 5.86] &99.9999 & 1.48 \\
\hline
Remote Control & 5 & 99.999 & 2 &  [1.62, 5.86] & 99.9999 &1.48 \\
\hline
Navigation systems& 10 & 99.99999 & 3 & [2.45, 6.80] &  99.99999 & 1.01 \\
\hline
Flight control systems& 10 & 99.9999999 & 3 & [3.85, 8.65] &  99.9999999 & 1.0 \\
\hline
\end{tabular}
\label{table2}
\end{table}
\section{Conclusion}
\label{Conclusion}
In this paper, we developed a reliability model for  URLLC/HRLLC networks  leveraging the  EC framework. The proposed model incorporates various latency components across the protocol stack, interactions between network (RLC, MAC and PHY) layers, particularly focusing on the acknowledged mode  of the RLC layer with separate TX and  RETX buffers. The model provides optimal  operating range for $\theta$ that satisfies URLLC constraints while maintaining  EC close to its maximum value, ensuring that the system can handle peak traffic loads with guaranteed  QoS. Our analysis reveals critical trade-offs between EC and reliability across different use cases offering valuable guidance for network design. This work serves as a valuable tool for system designers and network operators in optimizing the performance of mission-critical applications in Beyond-5G/6G networks. As future work, real-world implementation and validation of the proposed framework will be pursued to evaluate its practicality and effectiveness under realistic deployment scenarios.

\section*{Acknowledgement}
This work was funded by the Dutch "Rijksdienst voor Ondernemend Nederland – RVO" under the TSH number 21007 – RHIADA project. The authors would also like to acknowledge the project partners, GKN Fokker Aerospace and the Netherlands Aerospace Centre (NLR), for their support and collaboration. 

\bibliographystyle{IEEEtranUrldate}
\bibliography{references}
\end{document}